\documentclass[prd,showpacs,nofootinbib]{revtex4}
\usepackage{graphicx}
\usepackage{romannum}
\usepackage{xcolor}

\begin{document}

\title{Two-photon decay of P-wave positronium:  a tutorial}

\author{Abhijit Sen}
\affiliation{Novosibirsk State University, 630 090, Novosibirsk, Russia.}
\email{abhijit913@gmail.com}
\author{Zurab K.~Silagadze}
\affiliation{Novosibirsk State University  and Budker Institute of Nuclear
Physics, 630 090, Novosibirsk, Russia.}
\email{silagadze@inp.nsk.su}

\begin{abstract}
A detailed exposition of two-photon decays of P-wave positronium 
is given to fill an existing gap in the pedagogical literature.
Annihilation decay rates of P-wave positronium are negligible 
compared to the rates of radiative electric dipole transitions to the ground 
state. This circumstance makes such decays experimentally inaccessible.
However the situation is different for quarkonium and the experimental and 
theoretical research of two-photon and two-gluon decays of P-wave quarkonia 
is a still flourishing field.
\end{abstract}

\maketitle
\section{Introduction}
Two-photon decay rates of positronium in the P-state were calculated long
ago \cite{1,2}. A well-known textbook in quantum field theory \cite{3}
offers this problem as an exercise after presenting a basic tenets and
calculation tools of this theory.  

Being indeed an excellent exercise in quantum field theory, however we are
afraid that most students will find it too complicated. Even if they can find
the original papers about this problem \cite{1,2}, more modern presentation 
in \cite{4}, or its quarkonium counterpart in \cite{5}, this will not help
much, we think.    

This feeling is strengthened by the observation that in the unofficial 
solutions manual \cite{6} of the textbook \cite{3}, the decay rates of  
P-wave positronium are calculated incorrectly.

A detailed derivation of the two-photon amplitudes of various quarkonium
states can be found in \cite{7}. Although very useful, this paper uses
the Jacob-Wick helicity formalism  \cite{8}, not covered in any detail in
\cite{3} (however this formalism is briefly considered in older QFT textbook
\cite{9}), and therefore can seem somewhat esoteric for novices in quantum 
field  theory.

In this paper we attempt to fill this seeming gap in pedagogical literature
and provide a detailed calculation of the two-photon decay rates of P-wave
positronium along general style of the first five chapters of  \cite{3}.

The decay width of S-wave positronium in non-relativistic 
approximation can be obtained by elementary means \cite{9A}. Namely, 
the probability of electron-positron annihilation in S-wave positronium 
per unit time is
\begin{equation}
\Gamma=\rho v\sigma,
\label{eq1a}
\end{equation}
where $v$ is electron-positron relative velocity, $\rho=|\psi(0)|^2$ gives
a probability that the electron and positron meet each other in the 
positronium, and $\sigma$ is their annihilation cross-section when they meet.
The later can be related to the annihilation cross-section of the free 
electron-positron pair as follows. We must multiply the free cross-section 
by four, because it was averaged over the four possible spin-states of the 
incident electron and positron. Besides we must take into account the 
selection rules that only spin-singlet S-wave positronium can decay into 
two-photons, and only spin-triplet positronium can decay into three-photons. 
This selection rule can be enforced by taking $v\to 0$ limit which in the free 
cross-section leaves only s-wave contribution. Finally we must average over 
positronium polarization states which brings $1/(2J+1)$ factor in the formula. 
In this way we get the Pirenne-Wheeler formula \cite{32}:
\begin{equation}
\Gamma(Ps\to n\gamma)=\frac{1}{2J+1}\,|\psi(0)|^2\,\lim_{v\to 0} \left [
4v\sigma(e^+e^-\to n\gamma)\right ].
\label{eq1b}
\end{equation}
In the case of P-wave positronium the wave-function at the origin vanishes
and the decay amplitude becomes proportional to the spatial derivatives
of the wave function at the origin.  Correspondingly we need the free 
annihilation cross-section beyond the $v\to 0$ limit and 
things become much more complicated. As a result there is no Pirenne-Wheeler 
like simple way to get annihilation cross-section of P-wave positronium. 
The only thing which we can predict from the beginning is that in this case 
the annihilation rate will be suppressed compared to the S-wave positronium
annihilation rate by a factor $(|\vec{p}|/m)^2\sim \beta^2$ since this is 
a relative magnitude of the leading term in an expansion for small momenta 
\cite{9A}.

For light quarks the suppression goes away and the non-relativistic 
approximation breaks down completely. Even for heavy quarkonia, such as
charmonium, where $\beta^2\sim 0.3$, and bottonium, where $\beta^2\sim 0.1$, 
the relativistic corrections are important and these corrections
were studied in the frameworks of the Bethe-Salpeter equation \cite{10A,10B},
two-body Dirac equation \cite{10C}, covariant light-front approach \cite{10D},
sophisticated quarkonium potential model \cite{10E}, using an effective 
Lagrangian and QCD sum rules \cite{10F}, lattice QCD \cite{10G},  
non-relativistic QCD (NRQCD) \cite{10H}, to name a few. Regarding experimental
situation, see, for example, \cite{10I,10J}.

We hope that a detailed understanding of a more simple positronium case will
help students to navigate the vast literature devoted to the two-photon and
two-gluon decays of quarkonia.
 
\section{Positronium state vector}
A correct framework for relativistic bound state problem is provided by the
Bethe-Salpeter equation \cite{10,11} (for pedagogical discussions of this
equation see \cite{9,12}). Fortunately, for weakly bound non-relativistic
systems, like positronium, this notoriously difficult formalism simplifies 
considerably. It was shown \cite{13,14} that the relativistic two-fermion 
Bethe-Salpeter equation for such systems allows a systematic perturbation
theory and the corresponding lowest-order exactly solvable approximation 
essentially coincides to the Schr\"{o}dinger equation for a single effective 
particle.

At the lowest-order in fine structure constant $\alpha$, and in 
its rest frame, the positronium state vector can be approximated by the 
quantum state
\begin{equation}
|^{2S+1}L_J;M\hspace*{-1mm}>=\sqrt{2M_{Ps}}\int\frac{d\vec{p}}{(2\pi)^3}
\sum_{S_z=-S}^S \left[\begin{array}{ccc} l & S & J\\M-S_z & S_z & M\end{array}
\right]\tilde\psi_{lm}(\vec{p})|S,S_z\hspace*{-1mm}>,
\label{eq1}
\end{equation}
where $m=M-S_z$, 
$\tilde\psi_{lm}(\vec{p})=\int e^{-i\vec{p}\cdot\vec{x}}\psi_{lm}
(\vec{x})\,d\vec{x}$ is the momentum space  Schr\"{o}dinger wave
function of positronium (the principal quantum number is not indicated) giving 
the probability amplitude of finding the electron and positron with relative 
momentum $\vec{p}$ in the positronium, $M_{Ps}\approx 2m$ is the positronium 
mass ($m$ being the electron mass, not to be confused with the 
magnetic quantum number $m$ in $\tilde\psi_{lm}$) and the $\sqrt{2M_{Ps}}$ 
factor ensures a proper normalization of the positronium state vector 
consistent to the normalization of one-particle states adopted in \cite{3}. 
The Clebsch-Gordan coefficients 
$\left[\begin{array}{ccc} l & S & J\\m & S_z & M\end{array}\right]$
(a square bracket notation of \cite{15} is used for these coefficients)
couple the angular momentum eigenstates $\tilde\psi_{lm}(\vec{p})$ with
the total spin eigenstates $|S,S_z\hspace*{-1mm}>$ to form the total momentum
eigenstates $|^{2S+1}L_J;M\hspace*{-1mm}>$ . The total spin eigenstates by 
themselves are the result of quantum addition of electron and positron spins:
\begin{equation}
|S,S_z\hspace*{-1mm}>=\left[\begin{array}{ccc} \frac{1}{2} & \frac{1}{2} & 
S\\s &S_z-s  & S_z\end{array}\right]a^+(s,\vec{p})b^+(S_z-s,-\vec{p})
|0\hspace*{-1mm}>.
\label{eq2}
\end{equation}
Here $a^+(s,\vec{p})$ is the creation operator of electron with 
spin-projection $s$ and momentum $\vec{p}$, while $b^+(S_z-s,-\vec{p})$ is
the creation operator of positron with spin-projection $S_z-s$ and momentum 
$-\vec{p}$.
 
Note that, since particle number is not conserved in relativistic
quantum field theory, in general positronium state vector may contain 
contributions from Fock states that have particles other than ``valence'' 
electron and positron, as in (\ref{eq1}). However, in positronium, thanks to 
its non-relativistic nature, such admixtures are very small. For example, the 
the probability of finding  relativistic relative momenta, $p\sim m$ or 
higher, in positronium is only $O(\alpha^5)\sim 10^{-11}$ \cite{15A}.

We use standard spectroscopic notation in (\ref{eq1}). In particular, $L$
refers to the orbital angular momentum quantum number $l$ written as $S, P, D,
F,\ldots$ for $l=0,1,2,3,\ldots$

Electron and positron spins can combine to give either a total spin zero 
singlet state or a total spin one triplet states. The corresponding non-zero
Clebsch-Gordan coefficients are \cite{16}
\begin{eqnarray} &&
\left[\begin{array}{ccc} 1/2 & \;\;1/2 & 0 \\ 
1/2 & -1/2  & 0 \end{array}\right]=
-\left[\begin{array}{ccc} \;\;1/2 & 1/2 & 0 \\ 
-1/2 & 1/2  & 0 \end{array}\right]=\frac{1}{\sqrt{2}},
\;\;\left[\begin{array}{ccc} 1/2 & 1/2 & 1 \\ 
1/2 & 1/2  & 1 \end{array}\right]=1,\nonumber \\ && 
\left[\begin{array}{ccc} 1/2 & \;\;1/2 & 1 \\ 
1/2 & -1/2  & 0 \end{array}\right]= 
\left[\begin{array}{ccc} \;\;1/2 & 1/2 & 1 \\ 
-1/2 & 1/2  & 0 \end{array}\right]=\frac{1}{\sqrt{2}},\;\;
\left[\begin{array}{ccc} \;\;1/2 & \;\;1/2 & \;\;1 \\ 
-1/2 & -1/2  & -1 \end{array}\right]=1.
\label{eq3}
\end{eqnarray}
Using them, we easily get $S$-wave positronium state vectors
\begin{eqnarray} &&
|^1S_0;0\hspace*{-1mm}>=2\sqrt{m}\int\frac{d\vec{p}}{(2\pi)^3}\,
\tilde\psi_{00}(\vec{p})\,\frac{1}{\sqrt{2}}\left [a^+\left(\frac{1}{2},\vec{p}
\right )\,b^+\left(-\frac{1}{2},-\vec{p}\right)-a^+\left(-\frac{1}{2},\vec{p}
\right)\,b^+\left(\frac{1}{2},-\vec{p}\right)\right]|0\hspace*{-1mm}>, 
\nonumber \\ &&
|^3S_1;0\hspace*{-1mm}>=2\sqrt{m}\int\frac{d\vec{p}}{(2\pi)^3}\,
 \tilde\psi_{00}(\vec{p})\,\frac{1}{\sqrt{2}}\left [a^+\left(\frac{1}{2},
\vec{p}\right)\,b^+\left(-\frac{1}{2},-\vec{p}\right)+a^+\left(-\frac{1}{2},
\vec{p}\right)\,b^+\left(\frac{1}{2},-\vec{p}\right)\right]|0\hspace*{-1mm}>, 
\nonumber \\ &&
|^3S_1;1\hspace*{-1mm}>=2\sqrt{m}\int\frac{d\vec{p}}{(2\pi)^3}\,
\tilde\psi_{00}(\vec{p})\,a^+\left(\frac{1}{2},\vec{p}\right)\,b^+\left(
\frac{1}{2},-\vec{p}\right )|0\hspace*{-1mm}>,
\nonumber \\ &&
|^3S_1;-1\hspace*{-1mm}>=2\sqrt{m}\int\frac{d\vec{p}}{(2\pi)^3}\,
\tilde\psi_{00}(\vec{p})\,a^+\left(-\frac{1}{2},\vec{p}\right)\,
b^+\left(-\frac{1}{2},-\vec{p}\right )|0\hspace*{-1mm}>
\label{eq4}
\end{eqnarray}
Slightly abusing a notation (by using $s,s^\prime$ as matrix indexes) and 
changing the overall signs of some state vectors then necessary (in quantum 
theory state vectors are defined up to a phase), we can express (\ref{eq4}) 
state vectors in a more compact way:
\begin{equation}
|^{2S+1}S_J;M\hspace*{-1mm}>=2\sqrt{m}\int\frac{d\vec{p}}{(2\pi)^3}\,
\tilde\psi_{00}(\vec{p})\,\sum_{s\,s^\prime}a^+(s,\vec{p})[A^{(JM)}
(-i\sigma_2)]_{s\,s^\prime}b^+(s^\prime,-\vec{p}))|0\hspace*{-1mm}>,
\label{eq5}
\end{equation}
where $A^{(JM)}$ matrices are expressed through the Pauli matrices and the
triplet state polarization vectors
\begin{equation}
\vec{n}_1=\frac{1}{\sqrt{2}}\,(1,i,0),\;\; \vec{n}_{-1}=\frac{1}
{\sqrt{2}}\,(1,-i,0),\;\; \vec{n}_0=(0,0,1),
\label{eq6}
\end{equation}
in the following way
\begin{equation}
A^{(00)}=\frac{1}{\sqrt{2}},\;\;\;A^{(1M)}=\frac{1}{\sqrt{2}}\,
\vec{n}_M\cdot\vec{\sigma}.
\label{eq7}
\end{equation}

To deal with $P$-wave positronium  states, it is convenient instead of 
$\tilde\psi_{1m}(\vec{p})$ eigenstates  of the third component of the angular
momentum, to introduce Cartesian states
\begin{equation}
\tilde\psi^1(\vec{p})=\frac{1}{\sqrt{2}}\left(\tilde\psi_{1,-1}(\vec{p})-
\tilde\psi_{1,1}(\vec{p})\right),\;\;\;
\tilde\psi^2(\vec{p})=\frac{i}{\sqrt{2}}\left(\tilde\psi_{1,-1}(\vec{p})+
\tilde\psi_{1,1}(\vec{p})\right),\;\;\;\tilde\psi^3(\vec{p})=
\tilde\psi_{1,0}(\vec{p}).
\label{eq8}
\end{equation} 
We also will need the following non-zero $1\otimes 1$ Clebsch-Gordan 
coefficients \cite{16}:
\begin{eqnarray} &&
\left[\begin{array}{ccc} 1 & 1 & 1 \\ 1 & 0  & 1 \end{array}\right]=-
\left[\begin{array}{ccc} 1 & 1 & 1 \\ 0 & 1  & 1 \end{array}\right]=
\left[\begin{array}{ccc} 1 & \;\;1 & 1 \\ 1 & -1  & 0 \end{array}\right]=-
\left[\begin{array}{ccc} \;\;1 & 1 & 1 \\ -1 & 1  & 0 \end{array}\right]=
\left[\begin{array}{ccc} 1 & \;\;1 & \;\;1 \\ 0 & -1  & -1 \end{array}\right]=
-\left[\begin{array}{ccc} \;\;1 & 1 & \;\;1 \\ -1 & 0  & -1 \end{array}\right]
=\frac{1}{\sqrt{2}},\nonumber \\ && 
\left[\begin{array}{ccc} 1 & 1 & 2 \\ 1 & 0  & 1 \end{array}\right]=
\left[\begin{array}{ccc} 1 & 1 & 2 \\ 0 & 1  & 1 \end{array}\right]=
\left[\begin{array}{ccc} 1 & \;\;1 & \;\;2 \\ 1 & -1  & -1 \end{array}\right]=
\left[\begin{array}{ccc} \;\;1 & 1 & \;\;2 \\ -1 & 0  & -1 \end{array}\right]=
\frac{1}{\sqrt{2}},\;\;
\left[\begin{array}{ccc} 1 & 1 & 2 \\ 1 & 1  & 2 \end{array}\right]=
\left[\begin{array}{ccc} \;\;1 & \;\;1 & \;\;2 \\ -1 & -1  & -2 \end{array}
\right]=1,\nonumber \\ &&
\left[\begin{array}{ccc} 1 & \;\;1 & 0 \\ 1 & -1  & 0 \end{array}\right]=
\left[\begin{array}{ccc} \;\;1 & 1 & 0 \\ -1 & 1  & 0 \end{array}\right]=-
\left[\begin{array}{ccc} 1 & 1 & 0 \\ 0 & 0  & 0 \end{array}\right]=
\frac{1}{\sqrt{3}},\;\;
\left[\begin{array}{ccc} 1 & \;\;1 & 2 \\ 1 & -1  & 0 \end{array}\right]=
\left[\begin{array}{ccc} \;\;1 & 1 & 2 \\ -1 & 1  & 0 \end{array}\right]=
\frac{1}{2}\left[\begin{array}{ccc} 1 & 1 & 2 \\ 0 & 0  & 0 \end{array}
\right]=\frac{1}{\sqrt{6}}.
\label{eq9}
\end{eqnarray}
Using (\ref{eq8}) and (\ref{eq9}), we get for the scalar $^3P_0$ state
\begin{eqnarray} &&
|^3P_0\hspace*{-1mm}>=\frac{2}{\sqrt{6}}\sqrt{m}\int\frac{d\vec{p}}{(2\pi)^3}\,
\left (\tilde\psi^1(\vec{p})\left[a^+\left(\frac{1}{2},\vec{p}\right)\,
b^+\left(\frac{1}{2},-\vec{p}\right)-a^+\left(-\frac{1}{2},\vec{p}\right)\,
b^+\left(-\frac{1}{2},-\vec{p}\right)\right]-
\right .  \nonumber \\ && \left . 
i\tilde\psi^2(\vec{p})\left[a^+\left(\frac{1}{2},\vec{p}\right)\,b^+\left(
\frac{1}{2},-\vec{p}\right)+a^+\left(-\frac{1}{2},\vec{p}\right)\,
b^+\left(-\frac{1}{2},-\vec{p}\right)\right]-
\right .  \nonumber \\ && \left . \tilde\psi^3(\vec{p})\left[
a^+\left(\frac{1}{2},\vec{p}\right)\,b^+\left(-\frac{1}{2},-\vec{p}\right)+
a^+\left(-\frac{1}{2},\vec{p}\right)\,b^+\left(\frac{1}{2},-\vec{p}\right)
\right]\right).
\label{eq10}
\end{eqnarray} 
Analogous expressions can be obtained easily for three $^3P_1$ vector states
and for five $^3P_2$ tensor states and it can be checked that all of them
has an equivalent compact expression (up to an overall phases) of the form 
\begin{equation}
|^3P_J;M\hspace*{-1mm}>=2\sqrt{m}\int\frac{d\vec{p}}{(2\pi)^3}\,
\sum_{i=1}^3\tilde\psi^i(\vec{p})\,\sum_{s\,s^\prime}a^+(s,\vec{p})
[B^{(JM)i}(-i\sigma_2)]_{s\,s^\prime}b^+(s^\prime,-\vec{p}))|0\hspace*{-1mm}>,
\label{eq11}
\end{equation} 
where
\begin{equation}
B^{(00)i}=\frac{1}{\sqrt{6}}\,\sigma^i,\;\;
B^{(1M)i}=\frac{1}{2}\,\epsilon^{ijk}n_M^j\sigma^k,\;\;
B^{(2M)i}=\frac{1}{\sqrt{2}}\,h_M^{ij}\sigma^j.
\label{eq12}
\end{equation}
Here $h_M^{ij}$ are the polarization tensors for the $J=2,J_z=M$ states and it
is possible to construct them from the $\vec{n}_M$ polarization vectors 
\cite{17}: 
\begin{equation}
h_{\pm 2}^{ij}=n_{\pm 1}^in_{\pm 1}^j,\;\;h_{\pm 1}^{ij}=\frac{1}{\sqrt{2}}
\left(n_{\pm 1}^in_0^j+n_0^in_{\pm 1}^j \right),\;\;h_0^{ij}=\frac{1}
{\sqrt{6}}\left( n_1^in_{-1}^j+n_{-1}^in_1^j-2n_0^in_0^j\right).
\label{eq13}
\end{equation}
These $h_M^{ij}$ polarization tensors are traceless, symmetric, mutually 
orthogonal, and normalized to one:
\begin{equation}
\sum_{i=1}^3h_M^{ii}=0,\;\;\;\sum_{i,j=1}^3h_M^{ij}(h_{M^\prime}^{ij})^*=
\delta_{MM^\prime}.
\label{eq14}
\end{equation}
Besides, $h_{-M}^{ij}=(h_M^{ij})^*$.

At last, for the remaining three $^1P_1$ vector states we find equally easily
\begin{equation}
|^1P_1;M\hspace*{-1mm}>=2\sqrt{m}\int\frac{d\vec{p}}{(2\pi)^3}\,
\sum_i\tilde\psi^i(\vec{p})\,\sum_{s\,s^\prime}a^+(s,\vec{p})[C^{(1M)i}
(-i\sigma_2)]_{s\,s^\prime}b^+(s^\prime,-\vec{p}))|0\hspace*{-1mm}>,
\label{eq15}
\end{equation} 
where
\begin{equation}
C^{(1M)i}=\frac{1}{\sqrt{2}}\,n_M^i.
\label{eq16}
\end{equation}
It is clear from (\ref{eq5}), (\ref{eq11}) and (\ref{eq15}) that the 
positronium $<2\gamma|^{2S+1}L_J\hspace*{-1mm}>$ decay amplitude can be
expressed through two-photon annihilation amplitude 
$M(e^-e^+\to 2\gamma)=<2\gamma|e^-(s,\vec{p})e^+(s,-\vec{p})\hspace*{-1mm}>$ 
of the free electron and positron. So our next task is to study this 
annihilation amplitude.

\section{Two-photon annihilation amplitude of free electron and positron} 
\begin{figure}[htp!]
\begin{center}
\includegraphics[width=0.6\textwidth]{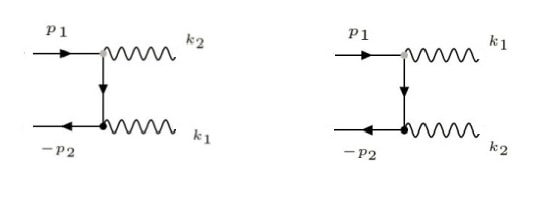}
\end{center}
\caption{Lowest order Feynman diagrams describing two-photon annihilation of 
free electron and positron.}
\label{Fig1}
\end{figure}

At the lowest order of the perturbation theory, $M(e^-e^+\to 2\gamma)$ is
described by two Feynman diagrams shown in Fig.~\ref{Fig1} and equals to
($e$ is electron's charge, $\epsilon_1\equiv\epsilon_1(k_1)$ and $\epsilon_2
\equiv\epsilon_2(k_2)$ are photon polarization vectors and we temporarily
suppress spin labels on spinors in this section)
\begin{equation}
M(e^-e^+\to 2\gamma)=-e^2\bar v(p_2)\left [\frac{\hat \epsilon^*_1
(\hat p_1-\hat k_1+m)\hat \epsilon^*_2}{(p_1-k_1)^2-m^2}+
\frac{\hat \epsilon^*_2(\hat p_1-\hat k_2+m)\hat \epsilon^*_1}
{(p_1-k_2)^2-m^2}\right ]u(p_1).
\label{eq17}
\end{equation}
Note that
$$(\hat p_1+m)\hat\epsilon^*\,u(p_1)=[2p_1\cdot\epsilon^*-\hat\epsilon^*
(\hat p-m)]u(p_1)=2p_1\cdot\epsilon^*u(p_1),$$
and (\ref{eq17}) is equivalent to
\begin{equation}
M(e^-e^+\to 2\gamma)=e^2\bar v(p_2)\left [\frac{2p_1\cdot\epsilon_1^*\,\hat
\epsilon_2^*-\hat\epsilon^*_2\hat k_1\hat\epsilon^*_1}{2p_1\cdot k_1}+
\frac{2p_1\cdot\epsilon_2^*\,\hat\epsilon_1^*-\hat\epsilon^*_1\hat k_2
\hat\epsilon^*_2}{2p_1\cdot k_2}\right ]u(p_1).
\label{eq18}
\end{equation}
But $p_1\cdot k_1\approx m^2-\vec{p}\cdot\vec{k_1}$, $p_1\cdot k_2\approx 
m^2+\vec{p}\cdot\vec{k_1}$, and $p_1\cdot k_1\,p_1\cdot k_2\approx m^4$. This
allows to rewrite (\ref{eq18}) as follows
\begin{eqnarray} &&
M(e^-e^+\to 2\gamma)=\frac{e^2}{2m^4}\bar v(p_2)\left [(m^2+\vec{p}\cdot
\vec{k}_1)(2p_1\cdot\epsilon_1^*\,\hat\epsilon_2^*-\hat\epsilon^*_2\hat 
k_1\hat\epsilon^*_1)+(m^2-\vec{p}\cdot\vec{k}_1)(2p_1\cdot\epsilon_2^*\,\hat
\epsilon_1^*-\hat\epsilon^*_1\hat k_2\hat\epsilon^*_2)\right]u(p_1)
\nonumber \\ && \approx
\frac{e^2}{2m^4}\bar v(p_2)\left (\begin{array}{cc} 0 & 2m^2a-m^2\bar{b}_++
\vec{p}\cdot\vec{k}_1\bar{b}_- \\ -2m^2a-m^2 b_++\vec{p}\cdot\vec{k}_1b_- & 0
\end{array}\right)u(p_1),
\label{eq19}
\end{eqnarray}
where
\begin{equation}
a=\vec{p}\cdot\vec{\epsilon}_1^{\,*}\,\vec{\sigma}\cdot\vec{\epsilon}_2^{\,*}+
\vec{p}\cdot\vec{\epsilon}_2^{\,*}\,\vec{\sigma}\cdot\vec{\epsilon}_1^{\,*},
\;\;
b_\pm=\vec{\epsilon}_1^{\,*}\cdot\vec{\sigma}\,k_2\cdot\sigma\,\vec{\epsilon}_2
^{\,*}\cdot\vec{\sigma}\pm\vec{\epsilon}_2^{\,*}\cdot\vec{\sigma}\,k_1\cdot
\sigma\,\vec{\epsilon}_1^{\,*}\cdot\vec{\sigma},\;\;\bar{b}_\pm=
\vec{\epsilon}_1
^{\,*}\cdot\vec{\sigma}\,k_2\cdot\bar{\sigma}\,\vec{\epsilon}_2^{\,*}\cdot
\vec{\sigma}\pm\vec{\epsilon}_2^{\,*}\cdot\vec{\sigma}\,k_1\cdot\bar{\sigma}\,
\vec{\epsilon}_1^{\,*}\cdot\vec{\sigma},
\label{eq20}
\end{equation}
and we have used the Coulomb gauge $\epsilon^\mu=(0,\vec{\epsilon})$ for photon
polarization vectors, and the chiral representation for gamma matrices.

It is convenient, following Alekseev \cite{2}, to express (\ref{eq19}) in 
terms of two-component Pauli spinors. In the chiral representation, adopted
in \cite{3},
\begin{equation}
u(p_1)=\left (\begin{array}{c}\sqrt{\sigma\cdot p_1}\,\xi \\
\sqrt{\bar{\sigma}\cdot p_1}\,\xi\end{array}\right )\approx \sqrt{m}\left (
\begin{array}{c}\left(1-\frac{\vec{\sigma}\cdot\vec{p}}{2m}\right )\xi
\\ \left(1+\frac{\vec{\sigma}\cdot\vec{p}}{2m}\right )\xi\end{array}
\right ),\;\; v(p_2)=\left (\begin{array}{c}\hspace*{3mm}
\sqrt{\sigma\cdot p_2}\,\zeta \\ -\sqrt{\bar{\sigma}\cdot p_2}\,
\zeta\end{array}\right )\approx \sqrt{m}\left ( \begin{array}{c}
\hspace*{3mm}\left(1+\frac{\vec{\sigma}\cdot\vec{p}}{2m}\right )\zeta
\\ -\left(1-\frac{\vec{\sigma}\cdot\vec{p}}{2m}\right )\zeta
\end{array}\right ),
\label{eq21}
\end{equation} 
where the approximate equalities, which are valid up to linear in $\vec{p}$
terms, follow from $\vec{p}_1=-\vec{p}_2=\vec{p}$ and
\begin{equation}
\sqrt{\sigma\cdot p}=\frac{\sigma\cdot p+m}{\sqrt{2(E+m)}}\approx \frac{1}
{2\sqrt{m}}\left(2m-\vec{\sigma}\cdot\vec{p}\right ),\;\;\;
\sqrt{\bar{\sigma}\cdot p}=\frac{\bar{\sigma}\cdot p+m}{\sqrt{2(E+m)}}\approx 
\frac{1}{2\sqrt{m}}\left(2m+\vec{\sigma}\cdot\vec{p}\right ).
\label{eq22}
\end{equation}
Substituting (\ref{eq21}) into (\ref{eq19}) and discarding quadratic and 
higher in $\vec{p}$ terms, we get
\begin{equation}
M(e^-e^+\to 2\gamma)\approx \frac{e^2}{2m^3}\,\zeta^+\left (-4m^2a+m^2(
\bar{b}_+-b_+)+\vec{p}\cdot\vec{k}_1(b_--\bar{b}_-)+\frac{m}{2}\,[b_++
\bar{b}_+,\,\vec{\sigma}\cdot\vec{p}\,]\right)\xi.
\label{eq23}
\end{equation} 
But, with required precision,
$$k_1\cdot (\sigma+\bar{\sigma})=k_2\cdot (\sigma+\bar{\sigma})=2m,\;\;
k_1\cdot (\sigma-\bar{\sigma})=-2\vec{k}_1\cdot\vec{\sigma},\;\;
k_2\cdot (\sigma-\bar{\sigma})=-2\vec{k}_2\cdot\vec{\sigma}=
2\vec{k}_1\cdot\vec{\sigma},$$
and therefore
\begin{equation}
b_\pm-\bar{b}_\pm=2(\vec{\epsilon}^{\,*}_1\cdot\vec{\sigma}\,\vec{k}_1\cdot
\vec{\sigma}\,\vec{\epsilon}^{\,*}_2\cdot\vec{\sigma}\mp \vec{\epsilon}^{\,*}_2
\cdot\vec{\sigma}\,\vec{k}_1\cdot\vec{\sigma}\,\vec{\epsilon}^{\,*}_1\cdot
\vec{\sigma}),\;\;b_++\bar{b}_+=2m\left(\vec{\epsilon}^{\,*}_1\cdot
\vec{\sigma}\,\vec{\epsilon}^{\,*}_2\cdot\vec{\sigma}+\vec{\epsilon}^{\,*}_2
\cdot\vec{\sigma}\,\vec{\epsilon}^{\,*}_1\cdot\vec{\sigma}\right)=
4m\,\vec{\epsilon}^{\,*}_1\cdot\vec{\epsilon}^{\,*}_2.
\label{eq24}
\end{equation}
Hence $[b_++\bar{b}_+,\,\vec{\sigma}\cdot\vec{p}\,]=0$. To simplify further,
we use the equality 
\begin{equation}
\vec{\sigma}\cdot\vec{a}\,\vec{\sigma}\cdot\vec{b}=
\vec{a}\cdot\vec{b}+i\vec{\sigma}\cdot(\vec{a}\times\vec{b}),
\label{eq24A}
\end{equation}
from which it follows that
\begin{equation}
\vec{\epsilon}^{\,*}_1\cdot\vec{\sigma}\,\vec{k}_1\cdot\vec{\sigma}\,
\vec{\epsilon}^{\,*}_2\cdot\vec{\sigma}=i\left[\vec{\epsilon}^{\,*}_2\cdot
(\vec{\epsilon}^{\,*}_1\times\vec{k}_1)+i\vec{\sigma}\cdot[(\vec{\epsilon}^
{\,*}_1\times\vec{k}_1)\times\vec{\epsilon}^{\,*}_2]\right ]=
-i\vec{k}_1\cdot(\vec{\epsilon}^{\,*}_1\times\vec{\epsilon}^{\,*}_2)-
\vec{\sigma}\cdot\vec{k}_1\,\vec{\epsilon}^{\,*}_1\cdot\vec{\epsilon}^{\,*}_2,
\label{eq25}
\end{equation}
because $\vec{k}_1\cdot\vec{\epsilon}^{\,*}_1=0$ and $\vec{k}_1\cdot\vec{
\epsilon}^{\,*}_2=-\vec{k}_2\cdot\vec{\epsilon}^{\,*}_2=0$. Using (\ref{eq25})
in (\ref{eq24}), we get
\begin{equation}
b_+-\bar{b}_+=-4i\,\vec{k}_1\cdot(\vec{\epsilon}^{\,*}_1\times
\vec{\epsilon}^{\,*}_2),\;\;\; b_--\bar{b}_-=-4\vec{\sigma}\cdot\vec{k}_1\,
\vec{\epsilon}^{\,*}_1\cdot\vec{\epsilon}^{\,*}_2,
\label{eq26}
\end{equation}
and the final form of the two-photon annihilation amplitude, valid up to
linear in $\vec{p}$ terms:
\begin{equation}
M(e^-e^+\to 2\gamma)\approx -\frac{2e^2}{m}\,\zeta^+\left[\vec{p}\cdot
\vec{\epsilon}^{\,*}_1\,\vec{\sigma}\cdot\vec{\epsilon}^{\,*}_2+\vec{p}\cdot
\vec{\epsilon}^{\,*}_2\,\vec{\sigma}\cdot\vec{\epsilon}^{\,*}_1
-i\,\vec{k}_1\cdot(\vec{\epsilon}^{\,*}_1\times
\vec{\epsilon}^{\,*}_2)+\frac{1}{m^2}\,\vec{p}\cdot\vec{k}_1\,\vec{\sigma}\cdot
\vec{k}_1\,\vec{\epsilon}^{\,*}_1\cdot\vec{\epsilon}^{\,*}_2\right]\xi.
\label{eq27}
\end{equation}
This result is consistent with the ones given in \cite{18} and  \cite{7} 
(after a typo is corrected in \cite{7} which lead to mutual interchange of the
photon polarization vectors).

\section{Two-photon decays of $S$-wave positronium}
To warm up, let's calculate two-photon decay width of $S$-wave positronium.
It is clear from (\ref{eq5}) that the decay amplitude has the following form
\begin{equation}
M(^{2S+1}S_J\to 2\gamma)=\frac{1}{\sqrt{m}}\int\frac{d\vec{p}}{(2\pi)^3}\,
\tilde\psi_{00}(\vec{p})\sum_{ss^\prime}[A^{(JM)}(-i\sigma_2)]_{ss^\prime}
M(e^-(s,\vec{p})\,e^+(s^\prime,-\vec{p})\to 2\gamma).
\label{eq28}
\end{equation}
In deriving prefactor in (\ref{eq28}), we have taken into account the 
relativistic normalization of the one-particle states $|e^-(s,\vec{p})
\hspace*{-1mm}>=\sqrt{2E_p}\,a^+(s,\vec{p})|0\hspace*{-1mm}>$ and that
$E_p\approx m$ at desired accuracy.

It follows from (\ref{eq27}) that
\begin{equation}
M(e^-(s,\vec{p})\,e^+(s,-\vec{p})\to 2\gamma)=\zeta^{s^\prime+}
\Lambda\xi^s,
\label{eq29}
\end{equation}
where $\Lambda$ is some $2\times 2$ matrix acting on spinor 
indices. But
\begin{equation}
\sum_{ss^\prime}[A^{(JM)}(-i\sigma_2)]_{ss^\prime}\zeta^{s^\prime+}
\Lambda\xi^s=
\mathrm{Tr}(\hat{\pi}_A^{(JM)}\Lambda),
\label{eq30}
\end{equation}
with
\begin{equation}
\hat{\pi}_A^{(JM)}=\sum_{ss^\prime}[A^{(JM)}(-i\sigma_2)]_{ss^\prime}
\xi^s\zeta^{s^\prime+}
\label{eq31}
\end{equation}
as a selector operator --- a $2\times 2$ matrix in spinor space which 
discriminates between the singlet $J=0$ and triplet $J=1$ states. 

To calculate $\hat{\pi}_A^{(JM)}$, let's recall that the particle and 
antiparticle two-component spinors are (note the flipped nature of 
antiparticle spinors) \cite{3}
\begin{equation}
\xi^\uparrow=\left (\begin{array}{c} 1 \\ 0 \end{array}\right ),\;\;
\xi^\downarrow=\left (\begin{array}{c} 0 \\ 1 \end{array}\right ),\;\;
\zeta^\uparrow=\xi^\downarrow=\left (\begin{array}{c} 0 \\ 1 \end{array}
\right ),\;\;
\zeta^\downarrow=-\xi^\uparrow=\left (\begin{array}{c} -1 \\ \hspace*{2mm} 0 
\end{array}\right).
\label{eq32}
\end{equation}
Then, because the only non-zero components of $-i\sigma_2$ are
$(-i\sigma_2)_{\uparrow\downarrow}=-1$ and $(-i\sigma_2)_{\downarrow\uparrow}
=1$, we will
have
$$\hat{\pi}_A^{(JM)}=\sum_{ss^\prime s^{\prime\prime}}A^{(JM)}_{ss^{\prime
\prime}}(-i\sigma_2)_
{s^{\prime\prime}s^\prime}\xi^s\zeta^{s^\prime+}=\sum_s\left(-A^{(JM)}_{s
\uparrow}\xi^s
\zeta^{\downarrow+}+A^{(JM)}_{s\downarrow}\xi^s\zeta^{\uparrow+}\right),$$
and
\begin{equation}
\hat{\pi}_A^{(JM)}=A^{(JM)}_{\uparrow\uparrow}\left (\begin{array}{c} 1 \\ 0 
\end{array}\right )(1,0)+A^{(JM)}_{\downarrow\uparrow}\left (\begin{array}{c} 
0 \\ 1 
\end{array}\right )(1,0)+A^{(JM)}_{\uparrow\downarrow}\left (\begin{array}{c} 
1 \\ 0 
\end{array}\right )(0,1)+A^{(JM)}_{\downarrow\downarrow}\left (\begin{array}
{c} 0 \\ 1 
\end{array}\right )(0,1)=\left(\begin{array}{cc} A^{(JM)}_{\uparrow\uparrow} &
A^{(JM)}_{\uparrow\downarrow}\\ A^{(JM)}_{\downarrow\uparrow} & A^{(JM)}_
{\downarrow\downarrow}
\end{array} \right).
\label{eq33}
\end{equation}
Note that $\hat{\pi}_A^{(JM)}$ and $A^{(JM)}$ act in different spaces --- the
first one acts on spinor indices and the second one acts on spin labels.
However (\ref{eq33}) indicates that in these spaces they act identically,
that is 
\begin{equation}
\hat{\pi}_A^{(00)}=\frac{1}{\sqrt{2}},\;\;\;
\hat{\pi}_A^{(1M)}=\frac{1}{\sqrt{2}}\,\vec{n}_M\cdot\vec{\sigma}.
\label{eq34}
\end{equation}
According to(\ref{eq27}), at zeroth order
\begin{equation}
\Lambda^{(0)}=\frac{2ie^2}{m}\,\vec{k}_1\cdot(\vec{\epsilon}^*_1
\times \vec{\epsilon}^*_2)
\label{eq35}
\end{equation}
doesn't depend on $\vec{p}$. Then the remaining integral
$$\int\frac{d\vec{p}}{(2\pi)^3}\,\tilde\psi_{00}(\vec{p})=\psi(\vec{x}=0)$$
just gives the position space positronium wave function at the origin.
Therefore 
\begin{equation}
M(^{2S+1}S_J\to 2\gamma)=\frac{2ie^2}{m\sqrt{m}}\, \vec{k}_1\cdot
(\vec{\epsilon}^*_1\times\vec{\epsilon}^*_2)\,\psi(\vec{x}=0)\,\mathrm Tr
(\hat{\pi}^{(JM)}_A). 
\label{eq36}
\end{equation}
To calculate the decay rate, we should module square the amplitude 
(\ref{eq36}), sum over the final state photon polarizations, average over the
initial state positronium polarizations, and integrate over the Lorentz
invariant final state phase space according to the general formula (overbar
indicates the above mentioned summation and averaging over polarizations,
$P=(M_{Ps},\vec{0})$ is the positronium 4-momentum)
\begin{equation}
d\Gamma(^{2S+1}L_j\to 2\gamma)=\frac{1}{2M_{Ps}}\,\overline{|M(^{2S+1}L_j\to 
2\gamma)|^2}\,\frac{d\vec{k}_1}{(2\pi)^32|\vec{k}_1|}\,\frac{d\vec{k}_2}
{(2\pi)^32|\vec{k}_2|}\,(2\pi)^4\delta^{(4)}(P-k_1-k_2).
\label{eq37}
\end{equation}
In Coulomb gauge, the photon polarization sums can be performed by using
\begin{equation}
\sum_{\epsilon}\epsilon^{*i}(\vec{k})\epsilon^{j}(\vec{k})=\delta^{ij}-
\frac{k^i\,k^j}{|\vec{k}|^2}.
\label{eq38}
\end{equation} 
Then (from here, it is assumed that repeated indices are implicitly summed 
over)
$$\sum_{\epsilon_1\,\epsilon_2}k_1^ik_1^j\epsilon^{imn}\epsilon^{im^\prime 
n^\prime}\epsilon_1^{*m}\epsilon_2^{*n}\epsilon_1^{m^\prime}\epsilon_2^
{n^\prime}=\epsilon^{imn}\epsilon^{jmn}k_1^ik_1^j=2|\vec{k}_1|^2,$$
and
\begin{equation}
\overline{|M(^{2S+1}S_j\to 2\gamma)|^2}=\frac{1}{2J+1}\sum_{\epsilon_1\,
\epsilon_2}|M(^{2S+1}S_j\to 2\gamma)|^2=\frac{8e^4}{m^3}\,|\vec{k}_1|^2\,
|\psi(\vec{x}=0)|^2\,|\mathrm Tr(\hat{\pi}^{(JM)}_A)|^2.
\label{eq39}
\end{equation} 
Since Pauli matrices are traceless and $\hat{\pi}^{(1M)}_A=\vec{n}_M\cdot
\vec{\sigma}/\sqrt{2}$, we immediately get that the spin-triplet $S$-wave 
positronium (orthopositronium) doesn't decay into two photons. Of course this
is just what is expected from the $C$-parity conservation in electromagnetic 
decays: for the two-photon final state $C=(-1)^2=1$ while for the $L=0,S=1$
orthopositronium $C=(-1)^{L+S}=-1$.

For the spin-singlet $S$-wave positronium (parapositronium) $\hat{\pi}^{(00)}
_A=1/\sqrt{2}$ and $\mathrm{Tr}(\hat{\pi}^{(00)}_A)=\sqrt{2}$. Then from 
(\ref{eq37}) and (\ref{eq39}) we get (the first $1/2$ factor accounts for the 
identity of the final state photons)
\begin{equation}
\Gamma(^1S_0\to 2\gamma)=\frac{1}{2}\,\frac{e^4|\psi(\vec{x}=0)|^2}
{\pi m^4}\int_0^\infty |\vec{k}_1|^2\delta(2m-2|\vec{k}_1|)\,d|\vec{k}_1|=
\frac{e^4|\psi(\vec{x}=0)|^2}{4\pi m^2}=\frac{4\pi\alpha^2}{m^2}\,
|\psi(\vec{x}=0)|^2.
\label{eq40}
\end{equation}
What remains is to use
$$|\psi(\vec{x}=0)|^2=\frac{m^3\alpha^3}{8\pi},$$
valid for the positronium ground state (for radial excitations with the
principal quantum number $n$ this quantity is $n^3$ times less), and obtain
a well known result of Pirenne and Wheeler \cite{19,20}
\begin{equation}
\Gamma(^1S_0\to 2\gamma)=\frac{m\alpha^5}{2}.
\label{eq41}
\end{equation} 

\section{Two-photon decays of $P$-wave positronium}
Hydrogen-like wave function of positronium has the form \cite{21}
\begin{equation}
\psi_{nlm}(\vec{r})=\left[\frac{1}{2n}\left(\frac{2}{na_0}\right)^3
\frac{(n-l-1)!}{(n+l)!}\right]^{1/2}\left(\frac{2r}{na_0}\right)^le^{-r/na_0}
\,\mathrm{L}_{n-l-1}^{2l+1}\left(\frac{2r}{na_0}\right)Y_{lm}(\theta,\phi),
\label{eq42}
\end{equation}
where $a_0=2/(m\alpha)$ is the Bohr radius for positronium and
\begin{equation}
L_n^m(x)=(n+m)!\sum\limits_{k=0}^n\frac{(-1)^k}{k!(n-k)!(k+m)!}\,x^k
\label{eq43}
\end{equation}
are associated Laguerre polynomials \cite{22}.

Some words of caution is perhaps appropriate here. In the physical and
mathematical literature one encounters two commonly used definitions
of Laguerre and associated Laguerre polynomials. This constitutes a possible
source of confusion to many students \cite{21,23}. In this paper we adopt the 
conventions of  Arfken and Weber \cite{22}, but one should bear in mind that
conventions used can change from book to book. For example, Landau and 
Lifshits \cite{24} use different conventions (namely, that of Spiegel 
\cite{25}) that changes the normalization coefficient, as well as the lower 
index from $n-l-1$ to $n+l$. Griffiths \cite{26} follow Arfken and Weber when 
relating associated Laguerre polynomials to the ordinary Laguerre polynomials 
but uses different normalization in the definition of the latter, and this 
leads to different normalization coefficient than in (\ref{eq42}).

It is clear from (\ref{eq42}) that $\psi_{nlm}(\vec{0})=0$, if $l\ne 0$. 
Therefore the zeroth-order approximation of the previous section cannot
be applied in the case of $P$-wave positronium decay and here our work in
the pre-previous section pays off: as (\ref{eq27}) shows
\begin{equation}
\Lambda=\Lambda^{(0)}+p^j\Lambda^{(1)j},
\label{eq44}
\end{equation}
where $\Lambda^{(0)}$ is given by (\ref{eq35}) and it doesn't 
contribute to the $P$-wave positronium decay, while
\begin{equation}
\Lambda^{(1)j}=-\frac{2e^2}{m}\left[\vec{\sigma}\cdot
\vec{\epsilon}_2^{\,*}\,
\epsilon_1^{\,*j}+\vec{\sigma}\cdot\vec{\epsilon}_1^{\,*}\,\epsilon_2^{\,*j}+
\frac{k_1^j}{m^2}\,\vec{\sigma}\cdot\vec{k_1}\,\vec{\epsilon}_1^{\,*}\cdot
\vec{\epsilon}_2^{\,*}\right].
\label{eq45}
\end{equation}
Positronium in the $^1P_1$ state with $S=0, L=1$ cannot decay into two 
photons due to $C$-parity conservation. It is reassuring  that our formalism
confirms this: $M(^1P_1\to 2\gamma)\sim p^j\tilde{\psi}^i(\vec{p})\,
\mathrm{Tr}(\hat{\pi}_C^{(1M)i}\Lambda^{(1)j})=0$, because 
$\Lambda^{(1)j}$ is 
traceless and 
$$\hat{\pi}_C^{(1M)i}=\sum_{ss^\prime}[C^{(1M)i}(-i\sigma_2)]_{ss^\prime}\xi^s
\zeta^{s^\prime+}=\frac{1}{\sqrt{2}}\,n_M^i$$
is proportional to the unit matrix.

As for the $^3P_J$ states, from (\ref{eq11}) we have 
\begin{equation}
M(^3P_J\to 2\gamma)=\frac{1}{\sqrt{m}}\,\mathrm{Tr}(\hat{\pi}_B^{(JM)i}
\Lambda^{(1)j})\int\frac{d\vec{p}}{(2\pi)^3}\,p^j
\tilde{\psi}^i(\vec{p}),
\label{eq46}
\end{equation}
where
\begin{equation}
\hat{\pi}_B^{(JM)i}=\sum_{ss^\prime}[B^{(1M)i}(-i\sigma_2)]_{ss^\prime}\xi^s
\zeta^{s^\prime+}=\left\{\begin{array}{l} \frac{\sigma^i}{\sqrt{6}},\;\;
\mathrm{if}\;\; (J,M)=(0,0), \\ \frac{1}{2}\epsilon^{ijk}n_M^j\sigma^k,\;\;
\mathrm{if}\;\; (J,M)=(1,M), \\ \frac{1}{\sqrt{2}}\,h_M^{ij}\sigma^j,\;\;
\mathrm{if} \;\; (J,M)=(2,M).\end{array}\right .
\label{eq47}
\end{equation}
Let us first calculate the integral in (\ref{eq46}). We have
\begin{equation}
\int\frac{d\vec{p}}{(2\pi)^3}\,p^j\tilde{\psi}^i(\vec{p})=\left . -i\nabla^j
\psi^i (\vec{x})\right|_{\vec{x}=0}.
\label{eq48}
\end{equation}
Since
$$\frac{1}{\sqrt{2}}\left(Y_{1,-1}(\theta,\phi)-Y_{1,1}(\theta,\phi)\right)=
\sqrt{\frac{3}{4\pi}}\,\frac{x}{r},\;\;\frac{i}{\sqrt{2}}\left(Y_{1,-1}(\theta,
\phi)+Y_{1,1}(\theta,\phi)\right)=\sqrt{\frac{3}{4\pi}}\,\frac{y}{r},\;\;
Y_{1,0}(\theta,\phi)=\sqrt{\frac{3}{4\pi}}\,\frac{z}{r},$$
it follows from (\ref{eq8}) and (\ref{eq42}) that Cartesian wave function 
of the $P$-wave positronium is 
\begin{equation}
\psi^i(\vec{x})=x^i\,\sqrt{\frac{3}{4\pi}\frac{1}{2n}\left(\frac{2}{na_0}
\right)^3\frac{(n-2)!}{(n+1)!}}\,\frac{2}{na_0}\,e^{-r/na_0}\,
\mathrm{L}_{n-2}^3
\left(\frac{2r}{na_0}\right),
\label{eq49}
\end{equation}
and therefore
\begin{equation}
\left . -i\nabla^j\psi^i (\vec{x})\right|_{\vec{x}=0}=-i\delta^{ij}\,
\sqrt{\frac{3}{4\pi}\frac{1}{2n}\left(\frac{2}{na_0}\right)^3\frac{1}
{(n-1)n(n+1)}}\,\frac{2}{na_0}\,\mathrm{L}_{n-2}^3(0).
\label{eq50}
\end{equation}
But from (\ref{eq43})
\begin{equation}
\mathrm{L}_{n-2}^3(0)=\frac{(n+1)!}{(n-2)!\,3!}.
\label{eq51}
\end{equation}
Combining  (\ref{eq48}), (\ref{eq50}) and (\ref{eq51}), we finally get
\begin{equation}
\int\frac{d\vec{p}}{(2\pi)^3}\,p^j\tilde{\psi}^i(\vec{p})=-i\delta^{ij}\,
\sqrt{\frac{n^2-1}{96\pi n^5}(m\alpha)^5},
\label{eq52}
\end{equation}
and
\begin{equation}
M(^3P_J\to 2\gamma)=-im^2\sqrt{\frac{(n^2-1)\alpha^5}{96\pi n^5}}\,\mathrm{Tr}
(\hat{\pi}_B^{(JM)i}\Lambda^{(1)i}).
\label{eq53}
\end{equation}
Now it's time to calculate traces in (\ref{eq53}) and this can be easily done
by using (\ref{eq24A}). The results are
\begin{equation}
\mathrm{Tr}(\hat{\pi}_B^{(00)i}\Lambda^{(1)i})=-2\sqrt{6}\,
\frac{e^2}{m}\,
\vec{\epsilon}^{\,*}_1\cdot\vec{\epsilon}^{\,*}_2,\;\;
\mathrm{Tr}(\hat{\pi}_B^{(2M)i}\Lambda^{(1)i})=-2\sqrt{2}\,
\frac{e^2}{m}\left(
2h_M^{ij}\epsilon^{\,*i}_1\epsilon^{\,*j}_2+\frac{\vec{\epsilon}^{\,*}_1\cdot
\vec{\epsilon}^{\,*}_2}{m^2}\,h_M^{ij}k_1^ik_1^j\right),
\label{eq54}
\end{equation}
and
\begin{equation}
\mathrm{Tr}(\hat{\pi}_B^{(1M)i}\Lambda^{(1)i})=-\frac{2e^2}{m}
\left[\vec{\epsilon}^{\,*}_2\cdot(\vec{\epsilon}^{\,*}_1\times\vec{n}_M)+
\vec{\epsilon}^{\,*}_1\cdot(\vec{\epsilon}^{\,*}_2\times\vec{n}_M)+
\frac{\vec{\epsilon}^{\,*}_1\cdot\vec{\epsilon}^{\,*}_2}{m^2}\,\vec{k}_1\cdot
(\vec{k}_1\times\vec{n}_M)\right]=0.
\label{eq55}
\end{equation}
The last result implies that $^3P_1$ state doesn't decay into two-photons
although it has positive $C$-parity. This time the decay is forbidden by
the Landau-Yang theorem \cite{27,28} which states that two real photons,
when referred to their center of mass frame, cannot be in a state of total 
angular momentum one. The proof of the theorem is simple and only makes use 
of such basic concepts as the superposition principle, Bose statistics and 
transversality of real photons, and rotational invariance \cite{29}. As we 
see, our formalism correctly reproduces this selection rule too.

In the case of $^3P_0$ state, the decay amplitude equals to
\begin{equation}
M(^3P_0\to 2\gamma)=i\frac{me^2}{2}\sqrt{\frac{(n^2-1)\alpha^5}{\pi n^5}}\,
\vec{\epsilon}^{\,*}_1\cdot\vec{\epsilon}^{\,*}_2.
\label{eq56}
\end{equation}
To calculate the decay rate, we should module square this amplitude and sum
over the photon polarizations using (\ref{eq38}). In this way we get
\begin{equation}
\sum_{\epsilon_1,\epsilon_2}|\vec{\epsilon}^{\,*}_1\cdot\vec{\epsilon}^
{\,*}_2|^2=
\sum_{\epsilon_1,\epsilon_2}\epsilon_1^{\,*i}\epsilon_1^{j}\,\epsilon_2^{\,*i}
\epsilon_2^{j}=2,
\label{eq56A}
\end{equation}
and
\begin{equation}
\overline{|M(^3P_0\to 2\gamma)|^2}=\sum_{\epsilon_1,\epsilon_2}
|M(^3P_0\to 2\gamma)|^2=8\pi m^2\,\frac{n^2-1}{n^5}\,\alpha^7.
\label{eq57}
\end{equation}
Eq.(\ref{eq37}) indicates that the decay rate and the corresponding squared 
amplitude are related by
\begin{equation}
\frac{d\Gamma}{d\Omega}=\frac{1}{64\pi^2M_{Ps}}\,\overline{|M|^2}.
\label{eq58}
\end{equation}
In our case the squared amplitude (\ref{eq57}) is just a constant,
the integration of (\ref{eq58}) is trivial and we finally get (the factor 
$1/2$ accounts for the identical final state photons)
\begin{equation}
\Gamma(^3P_0\to 2\gamma)=\frac{1}{2}\int \frac{d\Gamma(^3P_0\to 2\gamma)}
{d\Omega}\,d\Omega=\frac{n^2-1}{8n^5}\,m\alpha^7.
\label{eq59}
\end{equation}
The amplitude for the $^3P_2\to 2\gamma$ decay is more complicated: from
(\ref{eq53}) and (\ref{eq54}) we get
\begin{equation}
M(^3P_2\to 2\gamma)=ie^2m\sqrt{\frac{(n^2-1)\alpha^5}{12\pi n^5}}\left (
2h_M^{ij}\epsilon^{\,*i}_1\epsilon^{\,*j}_2+\frac{\vec{\epsilon}^{\,*}_1\cdot
\vec{\epsilon}^{\,*}_2}{m^2}\,h_M^{ij}k_1^ik_1^j\right).
\label{eq60}
\end{equation}
The next step is to module square this amplitude, perform the photon 
polarization sums, and average over the initial state positronium polarization
(that is sum over $M$ and divide by $2J+1=5$). This requires some algebra, 
somewhat simplified by the fact that $h_M^{ij}$ polarization tensors are 
traceless, symmetric and normalized to one. Besides, $\vec{k}_1^{\,2}\approx 
m^2$. After the dust settles we find
\begin{equation}
\overline{|M(^3P_2\to 2\gamma)|^2}=\frac{8\pi(n^2-1)m^2\alpha^7}{15n^5}\sum_M
\left(2-4h_M^{ij}h_M^{\,*im}\,\frac{k_1^jk_1^m}{m^2}+h_M^{ij}h_M^{\,*mn}\,
\frac{k_1^ik_1^jk_1^mk_1^n}{m^4}\right).
\label{eq61}
\end{equation}
Then from (\ref{eq58}) we get
\begin{equation}
\Gamma(^3P_2\to 2\gamma)=\frac{m\alpha^7}{32\pi}\,\frac{n^2-1}{15n^5}\sum_M
\left (8\pi-4h_M^{ij}h_M^{\,*im}\int\frac{k_1^jk_1^m}{m^2}\,d\Omega+
h_M^{ij}h_M^{\,*mn}\int\frac{k_1^ik_1^jk_1^mk_1^n}{m^4}\,d\Omega \right).
\label{eq62}
\end{equation}
The integrals involved give the completely symmetric second and forth rank
tensors respectively. As the integrands don't contain any external vector or
tensor, these tensors should be expressible in terms of $\delta^{ij}$ tensors 
only. Thus (the last expression is obtained by symmetrization of 
$\delta^{ij}\delta^{mn}$)
$$\int\frac{k_1^jk_1^m}{m^2}\,d\Omega=A\delta^{jm},\;\;\;
\int\frac{k_1^ik_1^jk_1^mk_1^n}{m^4}\,d\Omega=B(\delta^{ij}\delta^{mn}+
\delta^{im}\delta^{jn}+\delta^{in}\delta^{jm}).$$
To find the unknown coefficients $A$ and $B$, we simply contract these 
tensors:
$$4\pi=\int\frac{k_1^jk_1^j}{m^2}\,d\Omega=A\delta^{jj}=3A,\;\;\;
4\pi=\int\frac{k_1^ik_1^ik_1^mk_1^m}{m^4}\,d\Omega=B(\delta^{ii}\delta^{mm}+
\delta^{im}\delta^{im}+\delta^{im}\delta^{im})=15B.$$
Therefore $A=4\pi/3$, $B=4\pi/15$ and
\begin{equation}
\int\frac{k_1^jk_1^m}{m^2}\,d\Omega=\frac{4\pi}{3}\,\delta^{jm},\;\;\;
\int\frac{k_1^ik_1^jk_1^mk_1^n}{m^4}\,d\Omega=\frac{4\pi}{15}\,(\delta^{ij}
\delta^{mn}+\delta^{im}\delta^{jn}+\delta^{in}\delta^{jm}).
\label{eq63}
\end{equation}
It remains to substitute (\ref{eq63}) integrals into (\ref{eq62}) and remember
that $h_M^{ij}$ tensors are traceless and normalized to one. As a result, we 
get
\begin{equation}
\Gamma(^3P_2\to 2\gamma)=\frac{m\alpha^7}{32\pi}\,\frac{n^2-1}{15n^5}\sum_M
\left (8\pi-\frac{16\pi}{3}+\frac{8\pi}{15} \right)=\frac{n^2-1}{30n^5}\,
m\alpha^7.
\label{eq64}
\end{equation}
Our final results (\ref{eq59}) and (\ref{eq64}) are precisely the ones obtained
by Tumanov \cite{1} and Alekseev \cite{2}. In particular, for the $n=2$ 
states the decay widths are
\begin{equation}
\Gamma(2\,^3P_0\to 2\gamma)=\frac{3}{256}\,m\alpha^7,\;\;\;
\Gamma(2\,^3P_2\to 2\gamma)=\frac{1}{320}\,m\alpha^7.
\label{eq65}
\end{equation}

\section{Concluding remarks}
We hope that this rather detailed presentation of two-photon decays of
$P$-wave positronium will be helpful for quantum field theory students.
The standard approach used in this note is lucid enough and well motivated.
However thoughtful students can feel a necessity in a more powerful and
complete framework.

Our main assumption was a factorization of the bound state dynamics from the 
annihilation process. However such an approach violates energy conservation:
electron and positron that annihilate are on-shell and thus their total 
energy is greater than $M_{Ps}$. Of course for non-relativistic systems,
like positronium, the difference is of the order of $\alpha^2$ and can be
neglected at lowest order. However thoughtful students can wonder how the 
off-shellness of constituents can be reintroduced perturbatively at higher
order calculations (an example can be found in \cite{29A}).

The ratio $\Gamma(2\,^3P_0\to 2\gamma)/\Gamma(2\,^3P_2\to 2\gamma)=15/4$, which
follows from the previous section, is often quoted in the context of 
quarkonium decays. However it should be beared in mind that for  
quarkonia relativistic corrections are important and can lead to significant
modification of this ratio \cite{18}.

Another potential source of concern is subtle consequences of gauge 
invariance. As was shown by Low \cite{30}, gauge invariance and analyticity
implies that the decay amplitude of neutral bosons vanishes in the soft photon 
limit. Since the standard factorization treats intermediate charged states 
as on-shell, emission of soft photons by this particles is accompanied by
well known infrared singularities. Although the standard treatment ensures
the cancellation of these infrared singularities, the amplitude, for example,
for the $Ps\to 3\gamma$ decay, being finite, doesn't vanish in the soft photon 
limit, in contradiction to the Low's theorem \cite{31}. Special efforts to
correctly account for binding energy corrections are needed to reinforce the
theorem \cite{31}.

We hope that the two-photon decays of $P$-wave positronium can serve for
attentive students as a vista to a vast new land called a bound state problem
in relativistic quantum field theory. As an initial guidebook into this
interesting land, we recommend relatively recent PhD thesis \cite{32}.
 
\section*{Acknowledgments}
The work is supported by the Ministry of Education and Science of the Russian 
Federation.


\begin{thebibliography}{99}
\bibitem{1}
K.~A.~Tumanov, Quantum electrodynamics in configuration representation.
\Romannum{5}: two-photon annihilation of positronium. 
Zh.~Eksp.\ Teor.\ Fiz.\  {\bf 25}, 385-392 (1953).

\bibitem{2}
A.~I.~Alekseev, Two-photon annihilation of positronium in the P-state.
Sov.\ Phys.\ JETP {\bf 34}, 826-830 (1958).

\bibitem{3}
M.~E.~Peskin and D.~V.~Schroeder, {\it An introduction to quantum field 
theory} (Perseus Books, Reading, Massachusetts, 1995).

\bibitem{4}
V.~A.~Novikov, L.~B.~Okun, M.~A.~Shifman, A.~I.~Vainshtein, M.~B.~Voloshin 
and V.~I.~Zakharov, Charmonium and Gluons: Basic Experimental Facts and 
Theoretical Introduction,
Phys.\ Rept.\  {\bf 41}, 1-133 (1978).

\bibitem{5}
R.~Barbieri, R.~Gatto and R.~K\"{o}gerler,
Calculation of the Annihilation Rate of P-Wave Quark - anti-Quark Bound 
States,
Phys.\ Lett.\  {\bf 60B}, 183-188 (1976).

\bibitem{6}
Zhong-Zhi Xianyu, A Complete Solution to Problems in ``An Introduction to 
Quantum Field Theory''by Peskin and Schroeder.
https://zzxianyu.files.wordpress.com/2017/01/peskin\_problems.pdf
(accessed july 6, 2018).

\bibitem{7}
S.~U.~Chung, 
Two-Photon Amplitudes of Quarkonia in Helicity Formalism.
BNL preprint BNL-QGS-02-011, 2002.

\bibitem{8}
M.~Jacob and G.~C.~Wick,
On the general theory of collisions for particles with spin,
Annals Phys.\  {\bf 7}, 404-428 (1959) 
[Annals Phys.\  {\bf 281}, 774-799 (2000)].

\bibitem{9}
C.~Itzykson and J.-B.~Zuber,
{\it Quantum Field Theory} (McGraw-Hill, New York, 1980).

\bibitem{9A}
J.~M.~Jauch and F.~Rohrlich,
{\it The theory of photons and electrons. The relativistic quantum field 
theory of charged particles with spin one-half}
(Springer-Verlag, Berlin, 1980).

\bibitem{32}
C.~Smith,
{\it Bound State Description in Quantum Electrodynamics and 
Chromodynamics}, PhD dissertation, Louvain-la-Neuve, 2002.
https://cp3.irmp.ucl.ac.be/upload/theses/phd/smith.pdf
(accessed july 6, 2018).

\bibitem{10A}
C.~R.~Munz,
Two photon decays of mesons in a relativistic quark model,
Nucl.\ Phys.\ A {\bf 609}, 364-376 (1996).

\bibitem{10B}
G.~L.~Wang,
Annihilation rate of heavy 0++ P-wave quarkonium in relativistic Salpeter 
method,
Phys.\ Lett.\ B {\bf 653}, 206-209 (2007).

\bibitem{10C}
H.~W.~Crater, C.~Y.~Wong and P.~Van Alstine,
Tests of two-body Dirac equation wave functions in the decays of quarkonium 
and positronium into two photons,
Phys.\ Rev.\ D {\bf 74}, 054028 (2006).

\bibitem{10D}
C.~W.~Hwang and R.~S.~Guo,
Two-photon and two-gluon decays of p-wave heavy quarkonium using a covariant 
light-front approach,
Phys.\ Rev.\ D {\bf 82}, 034021 (2010).

\bibitem{10E}
S.~N.~Gupta, J.~M.~Johnson and W.~W.~Repko,
Relativistic two photon and two gluon decay rates of heavy quarkonia,
Phys.\ Rev.\ D {\bf 54}, 2075-2080 (1996).

\bibitem{10F}
J.~P.~Lansberg and T.~N.~Pham,
Effective Lagrangian for Two-photon and Two-gluon Decays of P-wave Heavy 
Quarkonium chi(c0,2) and chi(b0,2) states,
Phys.\ Rev.\ D {\bf 79}, 094016 (2009).

\bibitem{10G}
G.~T.~Bodwin,
NRQCD: Fundamentals and applications to quarkonium decay and production,
Int.\ J.\ Mod.\ Phys.\ A {\bf 21}, 785-792 (2006).

\bibitem{10H}
J.~P.~Ma and Q.~Wang,
Corrections for two photon decays of chi(c0) and chi(c2) and color octet 
contributions,
Phys.\ Lett.\ B {\bf 537}, 233-240 (2002).

\bibitem{10I}
M.~Ambrogiani {\it et al.},
Study of the gamma gamma decays of the $\chi_{c2}\,(1\,^3 P_2)$ and $\chi_{c0} 
\,(1\,^3 P_0)$ charmonium resonances,
Phys.\ Rev.\ D {\bf 62}, 052002 (2000).

\bibitem{10J}
K.~M.~Ecklund {\it et al.} [CLEO Collaboration],
Two-Photon Widths of the chi(cJ) States of Charmonium,
Phys.\ Rev.\ D {\bf 78}, 091501 (2008).

\bibitem{10}
E.~E.~Salpeter and H.~A.~Bethe,
A Relativistic equation for bound state problems,
Phys.\ Rev.\  {\bf 84}, 1232-1242 (1951).

\bibitem{11}
N.~Nakanishi,
A General survey of the theory of the Bethe-Salpeter equation,
Prog.\ Theor.\ Phys.\ Suppl.\  {\bf 43}, 1-81 (1969).

\bibitem{12}
Z.~K.~Silagadze,
Wick-Cutkosky model: An Introduction,
hep-ph/9803307.

\bibitem{13}
R.~Barbieri and E.~Remiddi,
Solving the {Bethe-Salpeter} Equation for Positronium,
Nucl.\ Phys.\ B {\bf 141}, 413-422 (1978).

\bibitem{14}
W.~E.~Caswell and G.~P.~Lepage,
Reduction of the {Bethe-Salpeter} Equation to an Equivalent Schr\"{o}dinger 
Equation, With Applications,
Phys.\ Rev.\ A {\bf 18}, 810-819 (1978).

\bibitem{15}
V. Devanathan,
{\it Angular Momentum Techniques in Quantum Mechanics} (Kluwer Academic 
Publishers, New York, 2002).

\bibitem{15A}
T.~Kinoshita and G.~P.~Lepage,
Quantum electrodynamics for nonrelativistic systems and high precision 
determinations of alpha,
Adv.\ Ser.\ Direct.\ High Energy Phys.\  {\bf 7}, 81-91 (1990).

\bibitem{16}
C.~Patrignani {\it et al.} [Particle Data Group],
Review of Particle Physics,
Chin.\ Phys.\ C {\bf 40}, 100001 (2016), p. 557.

\bibitem{17}
T.~Han, J.~D.~Lykken and R.~J.~Zhang,
On Kaluza-Klein states from large extra dimensions,
Phys.\ Rev.\ D {\bf 59}, 105006 (1999).

\bibitem{18}
Z.~P.~Li, F.~E.~Close and T.~Barnes,
Relativistic effects in gamma gamma decays of P wave positronium and q anti-q 
systems,
Phys.\ Rev.\ D {\bf 43}, 2161-2170 (1991).

\bibitem{19}
S.~Berko and H.~N.~Pendleton,
Positronium,
Ann.\ Rev.\ Nucl.\ Part.\ Sci.\  {\bf 30}, 543-581 (1980).

\bibitem{20}
D.~B.~Cassidy,
Experimental progress in positronium laser physics,
Eur.\ Phys.\ J.\ D {\bf 72}, 53 (2018).

\bibitem{21}
C.~E.~Burkhardt and J.~J.~Leventhal,
{\it Foundations of Quantum Physics} (Springer, New York, 2008).

\bibitem{22}
G.~B.~Arfken and H.~J.~Weber, 
{\it Mathematical Methods for Physicists} (Harcourt, New York, 2001).

\bibitem{23}
C.~E.~Burkhardt and J.~J.~Leventhal,
{\it Topics in Atomic Physics} (Springer, New York, 2006)

\bibitem{24}
L.~D.~Landau and E.~M.~Lifshits,
{\it Quantum Mechanics: Non-Relativistic Theory} (Butterworth-Heinemann,
Oxford, 1992).

\bibitem{25}
M.~R.~Spiegel, 
{\it Mathematical Handbook of Formulas and Tables} (McGraw-Hill, New York, 
1998).

\bibitem{26}
D.~J.~Griffiths, 
{\it Introduction to Quantum Mechanics} (Prentice Hall, New York, 1995).

\bibitem{27}
L.~D.~Landau,
On the angular momentum of a system of two photons,
Dokl.\ Akad.\ Nauk Ser.\ Fiz.\  {\bf 60}, 207-209 (1948).

\bibitem{28}
C.~N.~Yang,
Selection Rules for the Dematerialization of a Particle Into Two Photons,
Phys.\ Rev.\  {\bf 77}, 242-245 (1950).

\bibitem{29}
O.~Nachtmann,
{\it Elementary Particle Physics: Concepts and Phenomena} (Springer-Verlag,
Berlin, 1990).

\bibitem{29A}
J.~Pestieau, C.~Smith and S.~Trine,
Positronium decay: Gauge invariance and analyticity,
Int.\ J.\ Mod.\ Phys.\ A {\bf 17}, 1355-1398 (2002).

\bibitem{30}
F.~E.~Low,
Bremsstrahlung of very low-energy quanta in elementary particle collisions,
Phys.\ Rev.\  {\bf 110}, 974-977 (1958).

\bibitem{31}
J.~Pestieau and C.~Smith,
Soft photon spectrum in orthopositronium and vector quarkonium decays,
Phys.\ Lett.\ B {\bf 524}, 395-399 (2002).

\end{thebibliography}
\end{document}